\documentclass[twocolumn]{aa}
\usepackage{graphicx}

\begin{document}

\title{On hydrodynamic shear turbulence in Keplerian disks: \\
via transient growth to bypass transition}

\author{G. D. Chagelishvili, \inst{1} J.-P. Zahn, \inst{2}
        A. G. Tevzadze, \inst{1} \and
        J. G. Lominadze \inst{1}
       }
\offprints{J.-P. Zahn}

\institute{ Center for Plasma Astrophysics, Abastumani
Astrophysical Observatory, Kazbegi 2a, 380060 Tbilisi, Georgia \\
\email{aleko@tevza.org}
            \and
            LUTH, Observatoire de Paris, F-92190 Meudon, France \\
            \email{jean-paul.zahn@obspm.fr} 
           }

\date{Received 12 November 2002; accepted 7 February 2003}

\abstract{ This paper deals with the problem of hydrodynamic shear
turbulence in non-magnetized Keplerian disks. Several papers have
appeared recently on the subject, on possible linear instabilities
which may be due to the presence of a stable stratification, or
caused by deviations from cylindrical rotation. Here we wish to
draw attention to another route to hydrodynamic turbulence, which
seems to be little known by the astrophysical community, but which
has been intensively discussed among fluid dynamicists during the
past decade. In this so-called {\em bypass} concept for the onset
of turbulence, perturbations undergo a transient growth and if
they have initially finite amplitude they may reach an amplitude
that is sufficiently large to allow positive feedback through
nonlinear interactions. This transient growth is linear in
nature, and thus it differs in principle from the well-known
nonlinear instability. We describe the type of perturbations that
according to this process are the most likely to lead to
turbulence, namely non-axisymmetric vortex mode perturbations in
the two dimensional limit. We show that the apparently inhibiting
action of the Coriolis force on the dynamics of such vortical
perturbations is substantially diminished due to the pressure
perturbations, contrary to current opinion. We stress the
similarity of the turbulent processes in Keplerian disks and in
Cartesian flows and conclude that the prevalent skepticism of the
astrophysical community on the occurrence of hydrodynamic shear
turbulence in such disks is not founded.  }

\authorrunning{Chagelishvili et al. }
\titlerunning{On hydrodynamic shear turbulence in Keplerian disks}

\maketitle


\abstract{ This paper deals with the problem of hydrodynamic shear
turbulence in non-magnetized Keplerian disks. Several papers have
appeared recently on the subject, on possible linear instabilities
which may be due to the presence of a stable stratification, or
caused by deviations from cylindrical rotation. Here we wish to
draw attention to another route to hydrodynamic turbulence, which
seems to be little known by the astrophysical community, but which
has been intensively discussed among fluid dynamicists during the
past decade. In this so-called {\em bypass} concept of onset of
turbulence, finite amplitude perturbations undergo a transient
growth and they may reach an amplitude that is sufficiently large
to allow positive feedback through nonlinear interactions. This
transient growth is linear in nature, and thus it differs in
principle from the well-known nonlinear instability. We describe
the type of perturbations that according to this process are the
most likely to lead to turbulence, namely non-axisymmetric
vortical/aperiodic perturbations in the two dimensional limit. We
show that the apparently inhibiting action of the Coriolis force
on the dynamics of such vortical perturbations is substantially
diminished due to the pressure perturbations, contrary to current
opinion. We stress the similarity of the turbulent processes in
Keplerian disks and Cartesian flows and conclude that the
prevalent skepticism of the astrophysical community on the
occurrence of hydrodynamic shear turbulence in Keplerian disks is
not founded. }

\section{Introduction}

The emergence of X-ray astronomy exposed the importance of
accretion phenomena in many astrophysical systems, such as binary
stars, quasars and active galactic nuclei.  The possible
mechanisms governing accretion have been schematically understood
during the very first years of investigations (Shakura \& Sunyaev
1973; Pringle 1981): the inward transport of matter and the
outward transport of angular momentum in accretion disks was
ascribed to a turbulent (anomalous) viscosity.  The consequent
development in understanding the physics of this phenomenon and
the causes of that turbulence has been irregular and has taken
considerable time. But substantial progress has been achieved in
the nineties with the discovery of a linear instability in
magnetized disks (Balbus \& Hawley 1991, 1992, 1998; Hawley \&
Balbus 1991, 1992; Hawley et al. 1995; Stone et al. 1996). In
contrast, the solution of the accretion problem in the
non-magnetized case has not yet reached sufficient maturity.
Moreover, the very occurrence of turbulence in non-magnetized
disks has been questioned by Balbus et al. (1996) and Balbus \&
Hawley (1998). The reason for this situation is that cylindrical
flows with Keplerian profile belong to the class of smooth shear
flows, {\it i.e.} which present no inflection point; it is well
known that these flows are spectrally stable, although they may
become turbulent in the laboratory. The explanation of this
behavior remains one of the fundamental problems of fluid
mechanics. But the situation has changed in recent years, with the
recognition that the so-called nonmodal approach provides an
adequate, and even optimal mathematical formalism for describing
the dynamics of perturbations in plane-parallel shear flows. This
led to the emergence of the {\it bypass concept} for the onset of
turbulence in spectrally stable flows. This concept has triggered
much interest among fluid dynamicists, but it had little impact so
far on the astrophysical community. That is why feel it necessary
to present the salient results of that novel approach in a journal
which is read by the astrophysicists, to demonstrate its relevance
for the dynamics of Keplerian disks, and to join a bibliography of
the subject.

Our approach parallels that of Longaretti (2002), who analyzed the
existing experimental and numerical results, and who also
insisted, as we shall do here, on the similarity of subcritical
shear turbulence in plane and rotating flows.

Our paper is organized as follows.
In Sec. 2  we outline the fundamental problems of fluid mechanics
bearing on the description of plane-parallel smooth shear flows.
We present the recent developments in shear flow analysis made by
the hydrodynamic community, and the novel bypass concept for the
onset of turbulence in shear flows.
In Sec. 3 we analyze the quadratic forms of the dynamical
equations and discuss the essential role of pressure
perturbations, which counteract the effect of the Coriolis force
in 2D flows.
In Sec. 4 we describe the transient growth of two dimensional
perturbations. This will allow us to stress the similarity between
the behavior of Keplerian and Cartesian flows, and the negligible
role of the Coriolis force in the 2D perturbation kinematics and
energetics.
Finally we conclude in Sec. 5 that the bypass concept provides a
plausible scenario leading to turbulence in astrophysical disks.
In the Appendix we present a simple sketch of the bypass scenario.

\section {From spectral decay to transient growth and bypass
transition}

\subsection{On the onset of turbulence in spectrally stable shear
flows}

From the theoretical viewpoint there are flows that are
spectrally stable at all Reynolds numbers (e.g. plane Couette or
pipe Hagen-Poiseuille flows), while some others become spectrally
unstable at high enough Reynolds numbers (e.g. plane Poiseuille
or Blasius flows). In the latter case the flow is characterized by
the critical Reynolds number that is the marginal value of this
parameter above which the spectral instability occurs. In this
sense the critical Reynolds number of the spectrally stable flow
is infinity.

Here we shall consider smooth shear flows (without inflection
point in the velocity profile), like in Keplerian disks. These are
linearly stable according to classical fluid mechanics. As shown
by Rayleigh (1880), the existence of an inflection point (more
precisely of a vorticity extremum, as pointed out by
Fj{\o}rt{\o}ft 1950) in the equilibrium velocity profile is a
necessary condition for the occurrence of a linear (spectral)
instability in hydrodynamic flows. Thus smooth shear flows (flows
without a vorticity extremum) are ``relaxed'' in this context:
they are spectrally stable, meaning that exponentially growing
solutions are absent. However, it is well known from laboratory
experiments and from numerical simulations that finite amplitude
perturbations may cause a transition from laminar to turbulent
state at moderate, less then critical Reynolds number.

This has led to the development of the concept of {\it nonlinear
instability} in hydrodynamics (cf. Bayly 1986; Bayly et al. 1988;
Herbert 1988; Orszag \& Kells 1980; Orszag \& Patera 1980, 1983).
Until about ten years ago, the predominant view of this
laminar-turbulent transition was centered around the slow linear
amplification of exponentially growing perturbations (the familiar
T-S waves), which modify the flow profile and thereby allow a
secondary instability, further nonlinearity and finally a
breakdown to turbulent flow. According to this concept of
nonlinear instability, the perturbations (and the turbulent state
itself) are energetically sustained by nonlinear processes. This
concept of nonlinear instability has been borrowed by
astrophysicists, who still use it to explain turbulent processes
in smooth astrophysical flows, where no spectrally unstable
solution is known, and in particular in Keplerian disks flows (see
Balbus \& Hawley 1998). However, there are subcritical transition
phenomena that can not be attributed to the nonlinear
instability.

During the last decade of the 20th century, another viewpoint
emerged in the hydrodynamic community on understanding the onset
of turbulence in spectrally stable shear flows, labeled as {\it
bypass transition} (cf. Boberg \& Brosa 1988; Butler \& Farrell
1992; Reddy \& Henningson 1993; Trefethen et al. 1993; 
Morkovin 1993; Gebhardt \& Grossmann 1994; Henningson \& Reddy
1994; Baggett et al. 1995; Waleffe 1997; Grossmann 2000;
Reshotko 2001; Chagelishvili et al. 2002; Chapman 2002;
Rempfer 2003). Although the bypass transition scenario involves
nonlinear interactions -- which intervene once the perturbations
have reached finite amplitude -- the dominant mechanism leading to
these large amplitudes appears to be linear. This concept of the
onset of turbulence is based on the {\it linear transient growth
of vortex mode (aperiodic) perturbations}. The potential for
the transient growth has been recognized for more then a century
(see Kelvin 1887, Orr 1907a,b). However only recently has
the importance of the phenomenon 
been better understood (cf. Moffatt 1967; Marcus \& Press 1977;
Gustavsson \& Hultgren 1980; Craik \& Criminale 1986; Farrell \&
Ioannou 1993; Reddy \& Henningson 1993; Chagelishvili et al.
1997).

The bypass concept implies that the perturbation energy extracted
from the basic flow by linear transient mechanisms causes the
increase of the total perturbation energy during the transition
process. The nonlinear terms are conservative and only
redistribute the energy produced by the linear mechanisms. (A
simple sketch of the bypass scenario in a Keplerian flow is given
in the Appendix.)

Thus, according to this concept, the transient growth of
perturbations ({\it i.e.} the linear process) is the key element
in the transition to turbulence in spectrally stable flows. The
importance  of this linear process has been stressed in the titles
of several seminal papers: Henningson \& Reddy (1994) ({\em ``On
the role of linear mechanisms in transition to turbulence''});
Baggett et al. (1995) ({\em ``A mostly linear model of transition
to turbulence''}) and Reshotko (2001) ({\em ``Transient growth: A
factor in bypass transition''}).

\subsection{On the mechanism of perturbation amplification
in spectrally stable shear flows}

The scheme of investigation of shear flow dynamics implies the
following steps: introduction of perturbations into a mean flow,
linearization of the governing equations, and description of the
dynamics of perturbations and flow using the solutions of the
initial value problem ({\it i.e.} following temporal balances in
the flow). In principle this can be done, but in practice it is a
formidable task. Therefore the mathematical approach was changed.
This was done by assuming that the solution is separable in
eigenmodes, and then establishing the existence of at least one
unstable eigensolution. This approach became canonical in time,
and resulted in a shift of ground: attention was directed to the
asymptotic stability of the flow and {\it no attention} was paid
to any particular initial value or to the finite time period of
the dynamics. Indeed, this phase of the evolution was not thought
to have any significance -- it was left to speculation. But
recently the early transient period for the perturbations has been
shown to reveal ``rich'' and complicated behavior leading to
different consequences.

It was found in 1990s that smooth shear flows are crowded by
intense processes of mean flow energy extraction by perturbations,
energy exchange between perturbations, etc. even in the linear
approximation, while following the classical theory they are
spectrally stable and consequently ``relaxed.'' Especially it has
been shown that a superposition of decaying normal modes may grow
initially, but will eventually decay as time goes on -- a new {\it
linear} transient channel of energy exchange between the mean flow
and perturbations has appeared. Moreover, it has been shown that
transient growth can be significant even for subcritical values of
the Reynolds number and that its interplay with nonlinear
processes can result in transition to turbulence without any
``nonlinear instability'' of the flows.

\subsection{On the nonorthogonality of linear operators and the
nonmodal approach}

In fact an exact resonance is not necessary to obtain the
transient growth of perturbation. This is the consequence of the
non-normal character of operators that describe the linear
dynamics of perturbations in flows (see, {\it e. g.} Reddy et al.
1993). The fact that the eigenfunctions of the linearized
Navier-Stokes equations are not orthogonal ({\it i.e.} the
operator is non-normal) is enough to allow for solutions that
exhibit transient growth, depending on the initial conditions,
before finally decaying (see Criminale \& Drazin 1990; Trefethen
et al. 1993). The mechanism of the transient growth is
essentially inviscid -- the operators are highly non-normal for
large Reynolds numbers and the transient growth is asymptotically
large in $Re$. (This fact is extremely important in the case of
Keplerian accretion disks, where the Reynolds number is literally
astronomical: $Re>10^{10} !$)

These developments provoked a change of paradigm in the study of
linear processes in the considered flows. It was the so-called
{\it nonmodal approach} that became extensively used and even
canonized in the 1990s on these grounds. As a result, substantial
progress in the understanding of the shear flow phenomena has been
achieved. The nonmodal analysis -- some modification of the
initial value problem -- implicates the change of independent
variables from a laboratory to a moving frame and the study of
temporal evolution of spatial Fourier harmonics (SFH) of
perturbations without any spectral expansion in time (see Sec.~4).

\section {The similarity between Keplerian flow and plane parallel
shear flow}

Since our aim is to elucidate the basic similarity of the dynamics
of plane shear flows and of Keplerian disks, we approximate the
Keplerian flow in the two dimensional (2D) limit by its tangent plane
parallel flow, while retaining the effect of rotation through the
Coriolis force. This model, which ignores the purely geometrical
complications, is known as the shearing sheet model; it has been
has been used for the study of the linear dynamics of both wave
and vortex mode 2D perturbations (cf. Goldreich \& Lynden-Bell
1965; Goldreich \& Tremaine 1978; Drury 1980; Nakagawa \& Sekiya
1992 for wave mode perturbations, and Lominadze et al. 1988;
Fridman 1989 and Ioannou \& Kakouris 2001 for vortex mode
perturbations).

The dynamical equations are written in the local co-moving
Cartesian co-ordinate system:

\begin{equation}
x \equiv r - r_0 ; ~~~~~ y \equiv r_0 (\phi - \Omega_0 t) , ~~~~~
\label{coord}
\end{equation}
where ($r, \phi$) are standard cylindrical co-ordinates and
$\Omega_0$ is the local angular velocity at $r=r_0$:
\begin{equation}
 \Omega(r)= \Omega_0 +  {\partial  \Omega  \over \partial r}
 (r-r_0)\equiv \Omega_0 +  A {x \over r_0} . ~~~~~
\label{omegaexpand}
\end{equation}
$ A \equiv (r \partial_r \Omega )|_{r=r_0} $  is the shear
parameter (in Keplerian disks $A = -3\Omega_0/2 < 0$).

For the present purpose, we shall consider only 2D perturbations,
independent on the axial coordinate $z$. The resulting dynamical
linear equations for the perturbations of the radial velocity
($u_r$), the azimuthal velocity ($u_\phi$) and the pressure ($p$)
take the form:
\begin{equation}
{\partial  u_r  \over \partial t} + Ax {\partial  u_r \over
\partial y} - 2\Omega_0 u_\phi = - {\partial p  \over \partial
x}+\nu \Delta u_r , \label{ur}
\end{equation}
\begin{equation}{\partial  u_\phi  \over \partial t} + Ax {\partial
u_\phi \over \partial y}+ 2\Omega_0 u_r + A u_r = - {\partial p
\over \partial y}+\nu \Delta u_\phi , \label{uphi}
\end{equation}
\begin{equation}
{\partial  u_r  \over \partial x} + {\partial  u_\phi \over
\partial y} = 0 . \label{cont}
\end{equation}

The incompressible limit is taken here to leave out wave mode
perturbations and to keep only the vortex mode perturbations,
which are the basic ingredient of the bypass scenario. In this
case the energy density of the perturbation depends only on its
kinematic characteristics and not on the thermodynamics, such as
pressure.


Pressure terms are also absent when the dynamical equations are
written in vorticity form (for ${\bf {\omega }} = \hbox{\rm
curl}\, {\bf u}$), taking the curl of eqs. (\ref{ur}, \ref{uphi}).
This is probably the reason why the pressure perturbations have
been often ignored when discussing the dynamics of shear flows.
However they play a very important role, even in the
incompressible case, because they are the mediators of momentum
exchange between fluid particles, which results in the transient
growth of the vortex mode. (The physics of the ``mediator
activity" of the pressure perturbations is described in detail in
Chagelishvili et al. 1993, 1996.) The importance of the pressure
terms is clearly seen in the following analysis.

We multiply eqs. (\ref{ur}) and (\ref{uphi}) respectively by $u_r$
and $u_\phi$, in order to put them in kinetic energy form, and
average them over a domain which is symmetrical in $x$. This
procedure is similar to that performed in papers by Balbus et al.
(1996) and Balbus \& Hawley (1998).
\begin{equation}
{\partial \over \partial t} \langle {u_r^2 \over 2} \rangle =
2\Omega_0{ \langle {u_r u_\phi}\rangle } - \langle {{u_r}{\partial
p \over \partial x}} \rangle - \nu {\langle |\nabla u_r|^2
\rangle},
\label{ker}
\end{equation}
\begin{equation}
{\partial \over \partial t} \langle {u_\phi^2 \over 2} \rangle = -
[{2 \Omega_0} +A ]{ \langle {u_r u_\phi} \rangle } - \langle
{{u_\phi } {\partial p \over \partial y}} \rangle - \nu \langle
|\nabla u_\phi|^2 \rangle .
\label{kephi}
\end{equation}
The source of instability is represented by the shear term $A {
\langle {u_r u_\phi} \rangle }$. But the Coriolis term $2\Omega_0{
\langle {u_r u_\phi}\rangle }$ is explicitly present too, and in
eq. (\ref{kephi}) it overbalances the source term, since $A=-3/2
\Omega_0$ in the Keplerian disk. It thus would seem at first sight
that the Coriolis force has a profound influence on the shear flow
stability, as it has been argued by Balbus \& Hawley (1996). But
this conclusion is contradicted by the fact that the Coriolis
terms disappear when summing eqs. (\ref{ker}) and (\ref{kephi}),
meaning that the growth rate of the total kinetic energy is
independent of the Coriolis force.

A more detailed analysis will show that the Coriolis terms may
be eliminated also on the level of the dynamical equations,
by a suitable renormalization of the pressure perturbation.
In this two-dimensional and incompressible case, the
perturbation velocity field derives from the stream
function $\psi$:
\begin{equation}
u_r = - {\partial \psi \over \partial y} ; ~~~~~ u_\phi =
{\partial \psi \over \partial x} . ~~~~~
\end{equation}
Renormalizing the pressure perturbation
\begin{equation}
p^R \equiv p - 2\Omega_0 \psi , ~~~~~
\end{equation}
eqs. \ref{ur}--\ref{cont}, \ref{ker} and \ref{kephi} can be
rewritten as follows:
\begin{equation}
{\partial  u_r  \over \partial t} + Ax {\partial  u_r \over
\partial y}
 = - {\partial p^R  \over \partial x} +\nu \Delta v_r ,
\end{equation}
\begin{equation}
{\partial  u_\phi  \over \partial t} + Ax {\partial  u_\phi \over
\partial y}
 + A u_r = - {\partial p^R  \over \partial y} +\nu \Delta v_\phi ,
\end{equation}
\begin{equation}
{\partial  u_r  \over \partial x} + {\partial  u_\phi \over
\partial y} = 0 .
\end{equation}

\begin{equation}
{\partial \over \partial t} \langle {u_r^2 \over 2} \rangle = -
\langle {{u_r}{\partial p^R \over \partial x}} \rangle - \nu
{\langle |\nabla u_r|^2 \rangle} ,
\end{equation}
\begin{equation}
{\partial \over \partial t} \langle {u_\phi^2 \over 2} \rangle = -
A { \langle {u_r u_\phi} \rangle } - \langle {{u_\phi} {\partial
p^R \over \partial y}} \rangle - \nu \langle |\nabla u_\phi|^2
\rangle .
\end{equation}

In Sec. 2 we outlined the bypass concept in
Cartesian/planar shear flows (a simple sketch of the bypass
scenario is presented in the Appendix). Since we wish to apply a
similar scenario to differentially rotating disks, we shall write
the Cartesian counterparts of the averaged equations for
comparison. If $~y$ is the streamwise variable, and $~x$ is the
shearwise variable as in the considered case (i.e., $~x$ is analog
to $~r$, and $~y$ to $~\phi$), then the equations are:
\begin{equation}
{\partial \over \partial t} \langle {u_x^2 \over 2} \rangle = -
\langle {{u_x}{\partial p^c \over \partial x}} \rangle - \nu
{\langle |\nabla u_x|^2 \rangle} ,
\label{cartr}
\end{equation}
\begin{equation}
{\partial \over \partial t} \langle {u_y^2 \over 2} \rangle = -
A^c { \langle {u_x u_y} \rangle } - \langle {{u_y } {\partial p^c
\over \partial y}} \rangle - \nu \langle |\nabla u_y|^2 \rangle ,
\label{cartphi}
\end{equation}
where $~A^c$ is the shear parameter of the Cartesian flow
$U_0^c(0, A^cx)$; $~u_x$,  $~u_y~$ and  $~p^c~$ are the
perturbations of velocity and pressure, respectively.
 If the flow is set up with $~U_0^c~$ decreasing with $x$ (i.e.
$~A^c<0$), then eq. (\ref{cartphi}) shows that steady flow is
marked by transport in the direction of increasing $x$, as in the
Keplerian disk.

These equations are identical to the normalized equations derived
above for the Keplerian case, which stresses the similarity of
these two flows. But they deal with spatially averaged quantities,
and one may wish to establish that the similarity is even more
profound, namely that the local kinematics and energy of 2D vortex
perturbations of plane and disk flows are the same in time and
space, starting from identical initial conditions.

\section{Transient growth in 2D}

Mainly for the purpose of illustrating the bypass mechanism,
we shall describe now the evolution in time of an
initial perturbation, in two dimensions.  We perform
a spatial Fourier transform of all relevant variables, as
shown here for the pressure fluctuation:
\begin{equation}
p (x,y,t) = \hat p (k_x(t), k_y,t) \exp \left[ {\rm i} k_x(t)
x + {\rm i}k_y y \right] .
\label{fourp}
\end{equation}
In a shearing flow, the perturbations cannot keep the form of a
simple wave, since the wave-number of each spatial Fourier
harmonics (SFH) depends on time  (see Criminale \& Drazin 1990 for
a rigorous mathematical interpretation). In our geometry
\begin{equation}
k_x(t) = k_x(0) -  A k_y t ,
\label{kxdrift}
\end{equation}
thus the wave-number of each SFH varies in time along the flow
shear: it  ``drifts'' in ${\bf k}$-space. Therefore $p$, in eq.
(\ref{fourp}) above, depends on time through $k_x(t)$.

Equations (\ref{cont}), (\ref{ker}) and (\ref{kephi}) take the
following form in Fourier space:
$$ {{\rm d} \over{\rm d} t}\hat
u_r(k_x(t),k_y,t) - 2\Omega {\hat u_\phi}(k_x(t),k_y,t) =
~~~~~~~~~~~~~~~~~~~~~~~~~~~~~~~~~~~~~~~~~~~~~~~~~~~~~~~
$$
\begin{equation}
~~~~~~~~~~~~~~~~~~~~~~~~~~~~~~~ = -{\rm i}k_x(t){\hat
p}(k_x(t),k_y,t) ,
\label{fourur}
\end{equation}
$$ {{\rm d} \over {\rm d} t}{\hat u_\phi}(k_x(t),k_y,t) + (2\Omega
+ A) {\hat u_r}(k_x(t),k_y,t) = ~~~~~~~~~~~~~~~~~~~~~~~~~~~~~$$
\begin{equation} ~~~~~~~~~~~~~~~~~~~~~~~~~~~~~~~ =
- {\rm i}k_y{\hat p}(k_x(t),k_y,t) ,
\label{fourphi}
\end{equation}
\begin{equation}
k_x(t){\hat u_r}(k_x(t),k_y,t) + k_y {\hat u_\phi}(k_x(t),k_y,t) =
0 .
\label{fourcont}
\end{equation}
For simplicity, we ignored here the viscous forces; their action
may be easily incorporated in the analysis, as we shall see below.

One readily shows that these equations
possess a time invariant
\begin{equation}
{\cal I} = k_y {\hat u_r}(t) - k_x(t) {\hat u_\phi}(t) ,
\end{equation}
which expresses the conservation of vorticity in Fourier space.
Making use of this invariant, we may write the solution of the
system (\ref{fourur}--\ref{fourcont}) as follows:
\begin{equation}
{\hat u_r}(k_x(t),k_y,t)= {{k_y} \over {k_x^2(t) + k_y^2}}{\cal I}
,
\end{equation}
\begin{equation}
{\hat u_\phi}(k_x(t),k_y,t)= - {{k_x(t)} \over {k_x^2(t) +
k_y^2}}{\cal I} ,
\end{equation}
\begin{equation}
{\hat p}(k_x(t),k_y,t)= {{\rm i} \left[ A k_y^2  + 2 \Omega_0(
k_x^2(t)+ k_y^2)\right] \over {\left[k_x^2(t) + k_y^2\right]^2}}
{\cal I}.
\end{equation}
We see that the perturbed quantities vary aperiodically in time
(as is natural of vortex mode perturbations), and that they
undergo transient amplification. Energy is exchanged between the
perturbations and the background flow, and that represents the
basis of the bypass transition scenario. It is important to note
that the kinematic characteristics of the perturbations do not
depend on the rotation rate $\Omega_0$ -- they are identical to
those of Cartesian shear flows. Only the pressure perturbation
depends on $\Omega_0$. Consequently, the energetics of the SFH is
identical to that of the Cartesian case:
$$ {\hat
u}^2(k_x(t),k_y,t) \equiv {\hat u_r}^2(k_x(t),k_y,t) + {\hat
u_\phi}^2(k_x(t),k_y,t) =~~~~~~~~~~ $$
\begin{equation} = {1 \over {k_x^2(t) + k_y^2}}{{\cal I}^2} =
{\hat u}^2(k_x(0),k_y,0){{k_x^2(0) + k_y^2} \over {k_x^2(t) +
k_y^2}} ~.~
\label{fourke}
\end{equation}
We see that the kinetic energy reaches maximum amplitude for $k_x(t)=0$.
Similarly, the renormalized pressure perturbations do not
involve $\Omega_0$:
$$ {\hat
p^R}(k_x(t),k_y,t) = {\hat p}(k_x(t),k_y,t) - 2\Omega_0 {\hat
\psi}(k_x(t),k_y,t) =~~~~~$$
\begin{equation}
= {{\rm i} A k_y^2   \over {[k_x^2(t) + k_y^2]^2}} {\cal I} ~.~
\end{equation}
This expression is identical to that describing the evolution of
the pressure perturbation in a Cartesian flow, which proves the
similarity of transient growth in both types of flows, in two
dimensions.

The above analysis has been done in the inviscid case. Accounting
for the viscous forces is a straightforward procedure (see Fridman
1989): it may be done by multiplying the obtained equations by the
factor $ \exp \left[- \nu  \int_0^t {\rm d} t^{\prime}
(k_x^2(t^{\prime}) + k_y^2) \right] $. For instance, eq.
(\ref{fourur}) and (\ref{fourke}) will read as follows:
\begin{equation}
{\hat u_r}(k_x(t),k_y,t)= { k_y {\cal I} \over k_x^2(t) + k_y^2}
~e^{- \nu  \int_0^t {\rm d} t^{\prime} [k_x^2(t^{\prime}) +
k_y^2]} ~ ~,
\end{equation}
\begin{equation}
\hat u^2(k_x(t),k_y,t) = {{\cal I}^2 \over k_x^2(t) + k_y^2} ~
e^{- 2\nu  \int_0^t {\rm d} t^{\prime} [k_x^2(t^{\prime}) +
k_y^2]} ~.
\end{equation}

To estimate the maximum amplification which is achieved in this
transient growth at $k_y^2 \ll k_x^2(0)$, we assume that the
largest perturbation wave-number corresponds to the Kolmogorov
dissipation scale: $|k_x(0)|\, r_0 \approx Re^{3/4}$, where the
Reynolds number is defined as $Re = \Omega_0 r_0^2 / \nu$. Then
\begin{equation}
{\hat u^2_{\rm max} \over {\hat u}^2(k_x(0),k_y,0)} \approx
{k_x^2(0) \over k_y^2} \approx \left({Re \over (k_y \,
r_0)^2}\right)^{3/2} ,
\end{equation}
which can reach a huge value in astrophysical disks.

\section {Discussion and conclusions}

We have seen that the concept of nonlinear instability in smooth
spectrally stable shear flows has undergone substantial revision
in the hydrodynamic community. Nowadays the concept of {\it the
bypass transition to turbulence} has become  favorite and is under
intensive development.

Let us summarize the main features of this concept:\\
 -- the onset of turbulence
and the turbulent state itself in smooth spectrally stable shear
flows is supported energetically by {\it the linear transient
growth of vortex mode perturbations} -- the key ingredient of the
turbulence is the vortex mode (eddy) perturbation, and the key
phenomenon is the linear transient growth of the perturbation;\\
-- nonlinear processes {\it do not contribute to any energy
growth}, but regenerate vortex mode perturbations that are able to
extract shear flow energy. Doing so, nonlinear processes only {\it
indirectly} favor the energy extraction by the vortex mode
perturbations;
\\ -- the non-orthogonal nature of the linearized
Navier-Stokes equations is the formal basis of the transient
growth;
\\ -- the non-orthogonal nature increases with
increasing Reynolds number; thus the operators are highly
non-normal for the huge Reynolds numbers of Keplerian disks
$(Re>10^{10})$ and the transient growth is asymptotically large
in $Re$.

In this paper we wanted to contribute to the revival of
hydrodynamic shear turbulence as a possible explanation for the
``anomalous'' viscosity in non-magnetized Keplerian accretion
disks. Specifically, we wished to draw the attention of
astrophysicists to the bypass concept.

We showed that in disk flows there exist vortex mode perturbations
that are similar to those which are held responsible for the onset
of turbulence in the Cartesian shear flow. The key point of our
analysis is the interpretation of the important role of the
pressure perturbations in the dynamical processes.

In fact, the kinematics and energetics of the vortex mode
perturbations are identical in the rotating disk and the Cartesian
shear flows in the 2D case. The Coriolis force only causes
deviation of the pressure perturbations from the Cartesian case.
By focusing on the epicyclic motions, and underestimating the
action of the pressure terms, one is led to the false conclusion
that the Coriolis force suppresses hydrodynamic turbulence in
Keplerian flows (Balbus et al. 1996; Balbus \& Hawley 1998).

This property of the pressure perturbations has been established
here in the 2D case. In the more general case of 3D perturbations,
the dynamics of vortex mode perturbations is somewhat more
complicated, as it will be shown in a forthcoming paper (Tevzadze
et al.). But the role of the pressure perturbations is still to
counteract the Coriolis force. In the 3D disk case, the
wave-number domain where the perturbation undergoes transient
growth is smaller in comparison to 3D Cartesian flows, but this is
compensated by the very large Reynolds number characterizing
astrophysical disks.

Other scenarios which may lead to hydrodynamic turbulence in
astrophysical disks have been presented recently: they invoke a
linear (spectral) instability arising from the stratification
perpendicular to the disk (Dubrulle et al. 2002) or due to
deviations from cylindrical rotation (Urpin 2002; Klahr \&
Bodenheimer 2002). Such instabilities may well compete with the
bypass mechanism presented here, and it is not possible to
conclude presently which is the best candidate for rendering
astrophysical disks turbulent.

In laboratory experiments the field narrows, because there is no
stratification in the fluid (other than imposed on purpose), and
there the bypass mechanism provides an attractive explanation for
the turbulence detected in flows which are linearly stable
(angular momentum increasing outwards). We refer to Couette-Taylor
experiments performed by Wendt (1933), Taylor (1936), Coles (1950)
and Van Atta (1966). Very recently such turbulence has been
observed also in rotation profiles  which share the properties of
Keplerian disks, namely with their angular velocity decreasing
outwards (Richard 2001).

Decisive conclusions about the self-sustenance of the turbulent
state need to be supported by numerical simulations. To our
knowledge, the simulations reported so far have failed to detect
hydrodynamic shear turbulence in rotating flows with angular
momentum increasing outwards. In a recent paper, Longaretti (2002)
has discussed the possible explanations of this negative result.
The main reason is probably the lack of spatial resolution, which
still prevents from reaching even the Reynolds numbers at which
turbulence is detected in the laboratory. At this stage, we hope
that the analysis presented here will help to build a fruitful
background for the nonlinear numerical stability analysis of
rotating astrophysical shear flows. In any way, the definite
answer will result from the convergence of theoretical, numerical
and experimental work.

\begin{acknowledgements}
This work is supported by the International Science and Technology
Center grant G-553. G.D.C. would like to acknowledge the
hospitality of Observatoire de Paris (LUTH). The authors would
like to thank the referee
for his helpful comments on
the early version of the paper.
\end{acknowledgements}


\appendix

\section{Bypass scenario of the transition to turbulence}

\begin{figure}
  \centering
   \includegraphics[width=7cm]{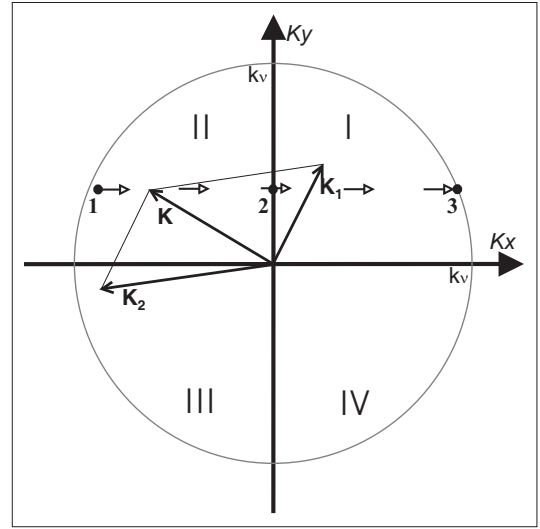}
   \caption{A simple sketch of the bypass scenario applied to
    Keplerian disk flow, in the wave-number plane ($k_x, k_y$).
    $k_x$ is the wave-number in the radial direction and
    $k_y$ in the azimuthal direction. The wave-number of the
    spatial Fourier harmonics (SPH) drifts from its initial
    position 1, the SPH is amplified in quadrant II, reaches
    maximum amplitude in 2, is attenuated in quadrant I,
    and undergoes viscous dissipation in 3. But the
    amplification quadrant II is repopulated through nonlinear
    interaction from SFHs located in the
    attenuation quadrants I and III. (See text for more details.)}
              \label{fig}
\end{figure}

We present here a simple sketch of the bypass scenario applied to
Keplerian disk flow, in the wave-number plane $~(k_x, k_y)$ (see
Fig.~\ref{fig}). We shall define as ``active domain'' the region
where viscous dissipation may be neglected, {\it i.e.} where
$~{k_x^2+k_y^2}<k_{\nu}^2$, with  $~k_{\nu}\approx
Re^{3/4}/{r_0}~$.

The linear dynamics of the perturbation may be described by
following each of its spatial Fourier harmonics (SFH) in this
wave-number plane. We single out one that is located initially at
some point 1 in the amplification part of the ``active domain'',
which meets the condition $~k_x/k_y < 0$. According to eq.
(\ref{kxdrift}), as $~k_x(t)~$ varies in time, the SFH drifts in
the direction marked by the arrows, since $~A<0$. (We present the
drift of the SFH only in the upper half-plane; since the
perturbation is real, there is a counterpart in the lower
half-plane.) Initially, as $~|k_x(t)|$ decreases, the energy of
the SFH grows. This growth lasts until the wave vector crosses the
line $~k_x=0~$ (point 2). Then, while undergoing attenuation, the
SFH continues its drift until it reaches point 3, where it is
dissipated through viscous friction. The same will occur with all
other Fourier harmonics. Consequently, if the nonlinear
interaction between different Fourier harmonics is inefficient,
the perturbation disappears eventually. Permanent extraction of
shear energy by the perturbations is necessary for their
maintenance, which is possible when quadrants II and IV where
$~k_x/k_y < 0$ are being repopulated through nonlinear interaction
between Fourier harmonics of quadrants I and III which have reached
sufficient amplitude. This may be achieved through three-wave
processes $~{\bf k_1} + {\bf k_2}\Rightarrow {\bf k}$; four wave
processes $~{\bf k_1} + {\bf k_2} + {\bf k_3}\Rightarrow {\bf k}$;
five wave processes, etc. In the figure we present an example of
three wave process $~{\bf k_1} + {\bf k_2}\Rightarrow {\bf k}~$
that contributes to the regeneration of an SFH in the
amplification area, transferring perturbation energy to it from
the attenuation areas. The bypass scenario implies the dominance
of this regeneration tendency of nonlinear processes, {\it i.e.}
predominant transfer of perturbation energy from quadrants I and
III to quadrants II and IV, in other words {\it nonlinear positive
feedback}. To what extent the reproduction of the Fourier
harmonics in quadrants II and IV is sustained, even in the case of
positive feedback, depends both on the amplitude and on the
spectrum of the initial perturbation. Nonlinear decay processes
are weak at low amplitudes and are not able to compensate the
linear drift of SFH in {\bf k}-plane. As a result, weak
perturbations are damped without any trace, and without inducing
transition to turbulence. The higher the amplitude of initial
perturbation, and the stronger are the nonlinear effects. At a
certain amplitude (which, of course, depends on the initial vortex
perturbation spectrum in {\bf k}-plane and on the Reynolds
number), the nonlinear processes are able to compensate the action
of the linear drift, thus ensuring permanent return of SFH to the
amplification areas (quadrants II and IV). This eventually ensures
a permanent extraction of energy from the background flow and
maintenance of the perturbations, and thus of turbulence.

Therefore the bypass scenario can be realized only in the case of
finite amplitude perturbations and in each case it has a threshold
that depends on the perturbation spectrum and the Reynolds number.

Let us summarize. According to the bypass concept, vortex mode
(eddy) perturbations are the basic ingredient of hydrodynamic
shear turbulence. The bypass scenario involves the interplay of
four basic phenomena:\\ -- the linear ``drift" of SFH in the {\bf
k}-plane;\\ -- the transient growth of SFH;\\ -- the usual viscous
dissipation;\\ -- nonlinear processes that close the feedback loop
of the transition by mixing -- by the angular redistribution of
SFH in {\bf k}-plane.

\end{document}